\documentclass[conference]{IEEEtran}
\IEEEoverridecommandlockouts
\usepackage{cite}
\usepackage{amsmath,amssymb,amsfonts}
\usepackage{algorithmic}
\usepackage{graphicx}
\usepackage{textcomp}
\usepackage{framed}
\usepackage{xcolor}
\usepackage{hyperref}
\usepackage{mathtools}
\hypersetup{
  colorlinks   = true,    
  urlcolor     = blue,    
  linkcolor    = blue,    
  citecolor    = red      
}
\usepackage[all]{hypcap}

\def\BibTeX{{\rm B\kern-.05em{\sc i\kern-.025em b}\kern-.08em
    T\kern-.1667em\lower.7ex\hbox{E}\kern-.125emX}}
\begin{document}

\colorlet{shadecolor}{yellow!35}

\title{Architecture of a Flexible and Cost-Effective Remote Code Execution Engine}

\author{
\IEEEauthorblockN{Ayaz Hafiz}
\IEEEauthorblockA{
\textit{Vanderbilt University}\\
Nashville, TN USA\\
ayaz.hafiz@vanderbilt.edu}
\and
\IEEEauthorblockN{Kevin Jin}
\IEEEauthorblockA{
\textit{Vanderbilt University}\\
Nashville, TN USA\\
kevin.jin@vanderbilt.edu}
}

\maketitle

\begin{abstract}

Oftentimes, there is a need to experiment with different programming languages and technologies when designing software applications. Such experiments must be reproducible and share-able within a team workplace, and manual effort should be minimized for setting up/tearing down said experiments.

This paper solves this problem by presenting a cloud-based web service for remote code execution, that is easily extensible to support any number of programming languages and libraries. The service provides a fast, reproducible solution for small software experiments and is amenable to collaboration in a workplace (via sharable permalinks). The service is designed as a distributed system to reliably support a large number of users, and efficiently manage cloud-hosting costs with predictive auto-scaling while minimizing SLA violations.
\end{abstract}


\section{Introduction}

\subsection{Motivation for Small Software Experiments}

In engineering workplaces, the most important part of software development
is software design. A large component of software design is the evaluation
and selection of software dependencies. In many cases, the best alternative
for a use case can only be chosen through experimentation of the alternatives'
integration with the use case. For example, one alternative may have a terse
API surface with strong performance guarantees, whereas another may have a
large API surface with looser guarantees; a software developer may wish to perform
some microbenchmarks and write sample integrations with both alternatives to
see which would be more prudent for the use case.

Setup and teardown of such experiments on an engineer's primary workspace may
be costly both in terms of time and resources. Installation of new programming
languages or technologies on a machine is time consuming and often induces a
rabbit hole of errors, like dependency mismatches or invalid environment
configurations which only take more time away from the engineer's primary workstreams.
Teardown is also costly and error-prone, as forgetting to remove one package
may eat up space on the workspace and wreck havoc when installing another
technology with a dependency on the long-ago-forgotten package (for example,
trying to build clang with an incomplete and outdated LLVM toolchain).

Sandboxed environments, like VMs or containers, are an option to avoid resource
waste and improve reproducibility, but take even more time and resources to setup.

Thus, there is a need for quickly-accessible, immediately-available services for
engineers to write and execute software experiments using a variety of technologies
without any additional setup on the user end.

\subsection{State-of-the-Art Services}

Several software services providing online code experimentation "playgrounds" are
widely available today. For example, Compiler Explorer\cite{CE} provides a suite of tools to write,
analyze, and execute C++ software with a variety of compilers and ecosystem packages.
However, Compiler Explorer\cite{CE} is heavily targeted toward C++ programs, and is
missing these desirable features for other programming languages. Try It Online\cite{TIO}
supports instant execution of software across a variety of programming languages, but
has no support for ecosystem packages. In the space of Cloud IDEs, players like
\emph{Github Code Spaces}\cite{GHCS} provide the "install and experiment" with any combination
of programming languages and technologies, but the setup cost of this is analogous to
that of a VM or container on a local workspace.

Thus, we see no existing services that cover all three desirable features of
supporting a variety of programming languages, supporting ecosystem packages in each
of those languages, and enabling instant execution of software without additional setup
from the user. Our contribution hopes to provide one such service.

\section{Design and Architecture}
\label{Design}

In this section, we discuss the architecture, design decisions made, and deep-dive concepts we used in the construction of the flexible remote code execution service (hereafter also referred to as just the ``service'').

\begin{figure}[htbp]
\centerline{\includegraphics[width=0.5\textwidth]{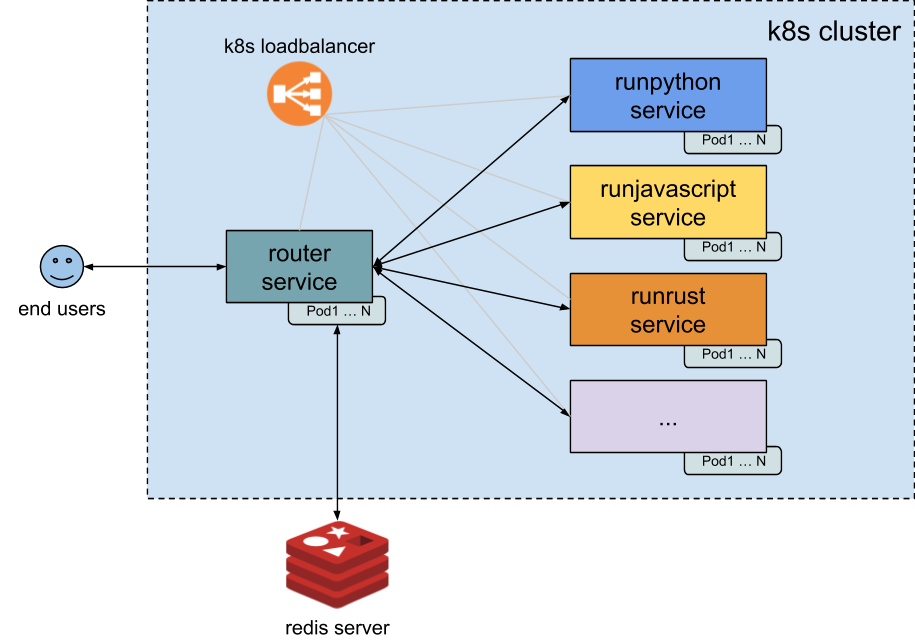}}
\caption{Request-level service architecture}
\label{req-level-arch}
\end{figure}

\subsection{Service Architecture}

The service consists of a set of front-end servers to serve web content and route code
execution requests and back-end servers
to execute user software in a sandbox containing a particular programming language(s) and
technologies.
Each front-end server is called a ``router'' and each back-end server is called ``runlang'',
where ``lang'' is replaced by the language ecosystem handled by a server. For example,
a server handling python code execution is called ``runpython''.

The set of front-end servers and each ``runlang'' service are exposed
as unique services in a cluster managed and load-balanced by Kubernetes. This enables
granular auto-scaling of the remote execution service as needed; for example, the number of
python code executors can be turned down without affecting the number of C++ code executors,
and the number of web servers can be turned up without affecting any code executors. This
approach has other advantages discussed later.

End users interact only with the service to front-end web servers, which supplies a playground
accessible in a web browser. When a user asks some code they have written in the playground
to be executed, the request is routed to a front-end web server via the Kubernetes load
balancer. This web server then creates and sends a code execution request to the appropriate
runlang service; for example runpython if the playground execution
request was for python code. This request arrives at a server in the
runlang service after hitting the service load balancer, and is executed
(possibly concurrently with other requests on the server) as a subprocess, and then the log output is propagated back the chain to the end user, when the playground UI reflects the execution results.

We now discuss the design of each component of the architecture in more detail.

\subsection{The router service}

As mentioned, the router service consists of a set of web servers, each called a ``router'',
that serves web content and routes code execution requests to back-end servers. The service
is managed by Kubernetes with Kubernetes auto-scaling and load balancing.

There are a few reasons for making the router service independent of code execution services:

\begin{itemize}
    \item The work done by each router server (validating HTTP requests, building and serving
    some HTML content from templates) is significantly less computationally expensive than the
    work done by each runlang server (actually executing user software
    of arbitrary complexity). By decoupling the "grunt work" of the web service from the more
    demanding parts, routers can be run on much cheaper machine resources than
    runlang servers.
    \item In general, the number of HTTP requests flowing to one router is going to be greater
    than the number of HTTP requests flowing to one code executor (as end users only interact
    with router servers). Decoupling code execution from the web server enables independent
    scaling of these two services, avoiding resource wastage.
\end{itemize}

We now discuss two additional features of the router service.

\subsubsection{Sharable playgrounds}

An interesting and important feature of the router web server is creation of sharable
playground URLs. This is a pivotal feature of our service, as we would like to support the sharing
of reproducible experiments between users in a workplace.

In general, there are two ways to encode client-side state in a URL - encode the state in the
URL itself and have logic on the client-side decode the state, or ask the server-side of the
application to store and load the state with some unique key. We chose the latter approach; while encoding
the playground state in URL parameters would avoid network calls and additional storage requirements
in the cloud, it would likely bring a poor user experience. The playground consists of text buffers
for a variety of different programming languages; to encode (ex. with base64 encoding) all the programming languages and the contents of their associated text buffers in a URL would produce a very large URL for the end user, not pleasant for sharing in most messaging applications.

\subsubsection{Authentication}

Since running software experiments can be computationally expensive and potentially dangerous, we
would like to avoid abuse of our service by bad actors. In an enterprise environment this may be
solved by hiding the service behind a private network, but since our service is on the Internet,
we solve this issue by registering a few privileged users and requiring authentication before
code execution requests can be made.

No authentication is needed for the runlang servers, as those are in a
private network only visible to the router service and all code execution requests coming
from router servers have already been authenticated.

\subsection{The runlang services}

As previously mentioned, each runlang service accepts and handles code execution for a particular
programming language/ecosystem configuration. For example, the runpython service executes code with
\texttt{cpython 3.7} and packages including \texttt{numpy} package; runrust executes code with 
\texttt{rustc nightly 11-15-20} and packages including \texttt{serde}. Users of the playground UI
choose which language/ecosystem configuration they would like to execute their code with, and
eventually that code is executed in a server in the service corresponding to that configuration.

In this section, we will describe the design of runlang servers and execution environments, and then
discuss some benefits to separating each runlang service.

\subsubsection{runlang server design}

Each runlang server provides a web server with RESTful APIs.
The web server used by all runlang service servers is identical so as to reduce the complexity
required in implementing the web servers. However, each server has runtime
knowledge of the programming language it is provisioned for and will always respond with an
error if it is asked to answer a query for a language/ecosystem configuration it is not provisioned
to do so. As discussed previously, the router is responsible for discovering runlang services
and routing user code execution requests to the appropriate service.

One API is \texttt{describe}, which provides information about the server's execution environment
(programming language, specification, compiler version, available packages) that the playground UI
uses to inform an end user of the environment their code will be run in and what features are
available to them.

\begin{figure}[htbp]
\centerline{\includegraphics[width=0.5\textwidth]{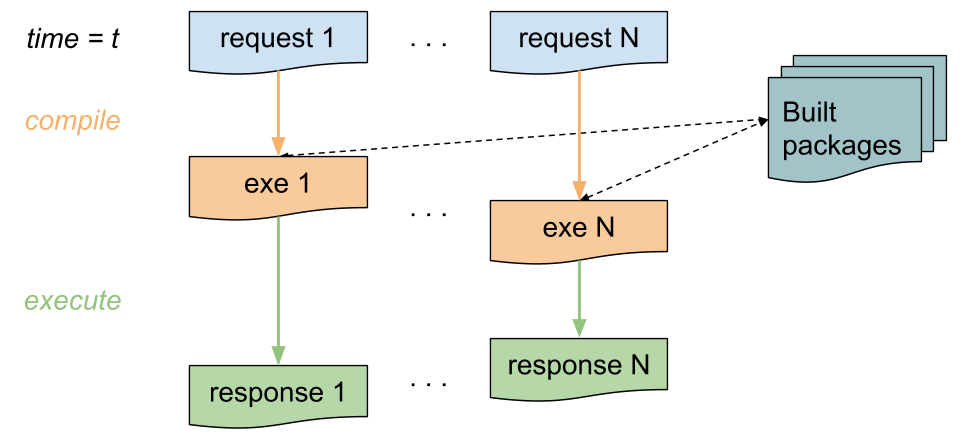}}
\caption{Model code executor pipeline, with the requirements of parallel request evaluation
and package build caching.}
\label{fig}
\end{figure}

The most interesting API is \texttt{execute}, which actually executes user code for a particular
language in the server execution environment. A handler for executing code given the server's configuration
(namely the language) is looked up, evaluated, and returned. While conceptually this process is very
simple, in practice trying to execute code with package dependencies is not so straightforward in some
languages, and strongly influences the design of the execution environments. This is the focus
of the next two sections.

\subsubsection{Requirements in executing user code}
\label{execreqs}

When executing user code in an environment where third-party software packages are available, we
would like to enforce the following requirements:

\begin{itemize}
    \item Execution should be concurrent; i.e. two separate code execution requests do not depend
    on each other, and their execution should not depend on each other.
    \item Execution should minimize re-compilation of software packages used in executed code as much
    as possible so as to provide fast response times and avoid resource waste. Ideally,
    ecosystem packages are built once and reused for all code execution requests.
\end{itemize}

The execution handlers must be implemented with these two requirements in mind, and the technical
challenges this presents are discussed in the \nameref{Implementation}.

\subsubsection{runlang execution environment images}

This brings us to the actual construction of the environment in which code for a language/ecosystem
configuration is executed. Each server in a runlang service is an instance of a Docker image for that
server configuration. For example, the runpython service consists of containers running ``runpython''
Docker images. Each runlang image is specialized for the particular language/ecosystem code it will
execute, but is derived from a common ``runlang\_base'' image that sets up the common web server
and API exposed by each instance.

To increase consistency between the build instructions (Dockerfiles) used among different runlang
images, reduce divergent configuration and errors between the base image and derived images, and
to more clearly reflect the langauge/ecosystem packages configuration of an image in a configuration
file, we designed a very small DSL named \texttt{imagegen} that produces artifacts needed for the build
and runtime of each runlang image. \texttt{imagegen} takes a YAML specification that describes the language
in the image, the packages in the image, how to install packages in the image, and how to validate
correctness of the image. \texttt{imagegen} validates that all required information for the runtime
of the image is provided, and generates a Dockerfile to build the image as well as some artifacts
to use in the image runtime (for example, descriptions of the image configuration used for the
\texttt{describe} API and for validation of execution requests to the server).

\subsubsection{runlang service separation}

As previously mentioned, one benefit of separating runlang services for different language/ecosystem
configurations is that it enables granular and independent auto-scaling of execution services.

Another large benefit is maintenance of individual language/ecosystem configurations. If one service
is responsible for execution of code in multiple language/ecosystems (say A and B), it is difficult
to modify A without also affecting B. By treating each configuration independently, debugging,
maintenance, and addition of new configurations becomes a very isolated and simple task.

\subsection{Virtual Private Cloud and Subnetting}
Most cloud providers like AWS, GCP and Azure support the concept of a VPC (Virtual Private Network), which can contain one or more subnets. A VPC can be thought of as a virtual network, with a router that manages routing between servers within and outside the network. The route tables for the router are configured on a subnet level. Subnets are essentially non-overlapping partitions of a VPC’s CIDR block. So for a VPC with a CIDR block of 10.0.0.0/18, possible subnets may include 10.0.0.0/24, 10.0.1.0/24, 10.0.2.0/22, etc. A public subnet is a subnet that has a route to the public Internet, while a private subnet is a subnet that has a route to a NAT gateway instead. This is why instances within a public subnet have public and private IP addresses, whereas in a private subnet they only have private IP addresses.

For an application to be accessible by general users over the Internet, it’s necessary that there is a public DNS that resolves to one or more public IP addresses for that application. Though, if the server hosting the application is in a public subnet, it poses a security risk as that server is more vulnerable to outside attacks. This is why for most production applications, the public DNS is actually mapped to the IP address of a proxy server in a public subnet, which then forwards the HTTP request to an application server within a private subnet.

\subsection{Infrastructure as Code}
Infrastructure as Code (IaC) is an important software engineering concept for several reasons. By codifying much of the state of the cloud environment, it’s possible to quickly re-setup any cloud resources in the situation that the cloud environment is lost. With all the configuration saved in the repository, it becomes much easier than trying to remember what cloud resources were originally there. Some large companies have thousands of different cloud services running together, and it’d be simply impossible to re-create the same environment from scratch if it was ever lost.

Another important benefit of IaC is that it allows engineers to understand the architecture of a cloud environment, even without direct access to an AWS, GCP or Azure account. Giving admin access to the cloud environment can lead to problems as engineers may mistakenly take down an application by de-provisioning the wrong server. Giving read-only access to a cloud environment can still be a security risk, since secrets are often stored on the cloud account and allowing all engineers to view those secrets increases the chances of a secret being leaked. These problems are exacerbated the larger a company is and the more engineers it has. If all of the cloud infrastructure is codified though, we could give engineers view access to the IaC configuration files, and allow them to understand the cloud infrastructure with minimal risk.

There are many publicly available tools to codify the cloud infrastructure for major cloud providers, including AWS, GCP and Azure, and also provision them automatically. Common IaC tools include Terraform and AWS Cloudformation, where cloud resources can be defined and a CLI tool can deploy/update/remove those resources in the cloud environment. These CLI tools can also be run by a continuous deployment pipeline, so that it’s not necessary for any specific engineer to be responsible for updating the cloud environment; the remote deploy agent can do that automatically.

\subsection{Continuous Deployment}
In a software development project with multiple engineers, continuous deployment (CD) can often accelerate the workflow of the entire team. This is because engineers would no longer have to worry about deploying the code they write to production servers, they can just focus on writing code instead. This is important because oftentimes, the process of deployment, depending on the complexity of the application, can be tedious and have multiple steps involved. If the deployment process was manual, it would be possible for an engineer to make a mistake during one of the steps, which could in the worst case, result in downtime of the application.

Another benefit of CD is that individual engineers would no longer require any secret tokens and keys in order to deploy their code to production. The only location that would require the secrets to deploy would be in the deploy agent itself, and secrets can usually be injected somewhere in the CD pipeline. By sharing secrets in fewer places, the attack surface is reduced, since the risk of those secrets possibly being leaked decreases.

\subsection{Predictive Horizontal Autoscaling}
Workloads can vary throughout the week and day, and can consume different amounts of system resources. It is important to allocate enough resources so that users are able to use the application without experiencing any service slowdowns, but not allocate too many resources, such that only a small percentage of resources is used at a high compute cost. Horizontal autoscaling is a method to scale up resources according to workload changes. A single or multiple metrics can be specified, such as CPU or memory usage, and if the autoscaler finds that servers in the cluster hit a specific metric threshold, then it can automatically provision additional resources. Likewise, an autoscaler can also determine if a cluster is under-utilizing system resources and decide to automatically scale down the number of servers. The most popular type of autoscaling is reactive autoscaling, which scales resources up and down according to real-time system metrics, however, the downside of this approach is two-fold:
\begin{enumerate}
\item since new servers are spun up retroactively, live requests cannot be re-routed to those new servers.
\item server provisioning is not instantaneous as there is a short delay until those servers can actually start handling traffic.
\end{enumerate}
These problems are amplified when the incoming traffic is more variable, since the autoscaler will constantly scale the resources up/down, trying to play catch-up with the real-time demand.
To mitigate the above problems, \emph{predictive} autoscaling was chosen to be used in this study instead. Predictive autoscaling forecasts the number of server replicas required in the future, so that those resources can be created beforehand to handle the incoming traffic. In this study, experiments have been conducted to evaluate the effectiveness of our predictive autoscaling implementation when dealing with a large, variable traffic workload.

\section{Implementation}
\label{Implementation}
We now discuss the implementation of each component in the service.

\subsection{Playground UI}

The playground UI is a simple multi-page web app using the \href{https://vuejs.org}{Vue} web framework.
The UI is stateful and communicates with the cloud service via standard HTTP requests.
Editor support in the playground is enabled via the \href{https://microsoft.github.io/monaco-editor/}{Monaco Editor},
which provides IDE-like editor support (syntax highlighting, completions) for a variety of programming
languages. The UI permits state-preserving cycling between buffers for different programming languages.

\subsection{AWS Elastic Kubernetes Service}
We used \href{https://aws.amazon.com/eks/}{AWS Elastic Kubernetes Service (EKS)} as a managed Kubernetes solution. The large benefit of using EKS over a self-hosted Kubernetes cluster is that it simplifies many steps required to create a cluster from scratch, such as choosing a CIDR block for pods in the cluster and deciding on a network policy like Calico or Flannel. EKS takes care of all of that for us, so we can just focus on writing the deployments, services and pods we want to add to our cluster. EKS also allows us to define how many servers we want to distribute the cluster across, and by changing just a single number in the configuration, EKS will appropriately spin up/down that many servers.

\subsection{The router service implementation}

As discussed in the \nameref{Design} section, the router service is a Kubernetes service consisting of a number
of router server pods across EC2 instances that are auto-scaled and load-balanced by the Kubernetes
runtime.

Each router pod exposes a RESTful API via a \href{https://flask.palletsprojects.com/en/1.1.x/}{Flask}
web server. The construction of the web server is uninteresting, but we would like to bring attention
to the implementation of two notable features mentioned in the router design.

\subsubsection{Sharable Playground IDs}

As mentioned in the \nameref{Design} section, the router service is responsible for generating  playground IDs
(\texttt{pg\_id}s), which enable sharing of a reproducible playground state between users
via a hyperlink. On the playground UI, when a user requests a shareable playground link, a
request for a \texttt{pg\_id} given the current state of the playground is sent to the router
service. A unique 10-digit hexadecimal \texttt{pg\_id} is generated and associated with the current playground code in a \href{http://redis.io}{Redis} cluster. The \texttt{pg\_id} is then used to create a hyperlink that is copied to the user's clipboard, and when accessed, will initialize the playground with that state.

We use Redis as an in-memory key-value store to ensure that saving and loading playground states
remains very fast for users. Furthermore, as a NoSQL database, Redis does not require a schema for how we represent the state of a playground, which permits flexibility in changing the representation over time.
While primarily being an in-memory service, Redis can also persist data (though with few
consistency guarantees); for our purposes at this time, an in-memory store is enough as we have
our Redis cluster managed by the highly-available \href{https://aws.amazon.com/elasticache/redis/}{AWS Elasticache}.

\texttt{pg\_id}s are limited to 10 hexadecimal digits to avoid the generation of overtly long
hyperlinks to playgrounds, which improves user experience in sharing hyperlinks. The space of 10-digit
hexadecimal IDs permits upwards of $10^{16}$ unique states, more than enough for our use case.

Currently, keys are stored in Redis with a TTL (time-to-live) of 30 days, before they are removed. This way older keys can be freed up for re-use.

\subsubsection{Authentication}

Authentication is supported by the \href{https://flask-login.readthedocs.io/en/latest/}{flask-login}
extension to the Flask web framework. As mentioned in the \nameref{Design} section, all publicly-exposed web APIs
are guarded by session authentication via a login form (flask-login uses cookies by default;
alternative options include JWTs and basic auth headers, which we avoided for simplicity
of integration).

Currently, registration of new users can only be done in an ``offline'' fashion; a database of
privileged login information is encrypted at rest and deployed with the router servers. Currently the
database cannot be modified at runtime, as we would like to avoid the additional complexity that comes
with automatically verifying new users (email conformation, bot detection, etc.). Moreover, in a
workplace environment where all potential users are known, dynamic registration is not necessary;
however, supporting such features would be a straightforward process. 

\subsection{The runlang service implementation}

As mentioned in the \nameref{Design}, each runlang service (runpython, runcpp, etc.) is deployed
as a Kubernetes service consisting of a number of runlang server pods for a particular language/ecosystem
configuration. As with the router service, each runlang service is auto-scaled and load-balanced
by the Kubernetes runtime.

Like the routers, each runlang instance exposes a web API via a Flask web server. Additionally, each runlang container validates at runtime that it can handle the code execution requests, using information it is instantiated with as part of the image build process.

The \texttt{imagegen} DSL used to generate Dockerfiles and build artifacts for runlang images is
written in Python. Implementation of the DSL as described in the design section is standard and uninteresting.

\subsubsection{Challenges in executing user code}

Recall the \hyperref[execreqs]{design requirements} of executing user code in runlang instances:
execution of independent requests should be concurrent, and dependencies should not be
re-built more than once (programs should be built in ``debug'' or low-optimization mode, where
whole-program optimization or other optimization tasks that would require rebuilding are not done).

For languages using compilers/interpreters with built-in global module systems, like Python or Node Javascript, the task of executing code with these requirements is as trivial as running code in that language directly - simply install the modules globally, and the language compiler/interpreter plays the role of a build system, discovering where those modules are without needing input from the invoker. Furthermore, each script can be run independently, concurrently, and the compiler/interpreter instances running those scripts will reuse the global modules without any additional work.

But for languages without built-in global module systems, like C++ or Rust, this task is not so simple.
One can build third-party packages and feed the build artifacts to compiler invocations themselves, or
use a package manager that generates this information for the compiler behind a much simpler interface
exposed to the user. Since the former approach, when used for multiple packages in a maintainable way
would eventually become a package manager itself, we decided to use an existing package manager to
handle ecosystem packages and builds in such languages.

However, package managers are often not \emph{just} package managers, but are also build systems. Such build
systems (namely, we used \texttt{Conan} for C++ and \texttt{cargo} for Rust) require information about
project structure and configuration, which can be problematic for the aforementioned requirements. If
a common project structure/configuration is enforced, then for two simultaneous execution requests A
and B, either separate projects must be forked off for A and B, which requires rebuilding ecosystem
packages (cache preservation is possible, but is error-prone and outside the scope of this work), or
``fill in'' the requests into the project and execute the project for each request sequentially, which
fails the requirement of concurrency.

All this is to say, executing software with third-party dependencies in such languages may become a
task of decoupling a package manager from a build system, discovering APIs in a build system to treat
source files as binaries rather than as part of a ``project'', or developing some hacks to accomplish
these goals. However, in our experience, we have been able to meet the desired requirements for all
currently supported languages without sacrificing machine resources or code complexity.

\subsection{AWS ElastiCache}
As mentioned earlier, we use AWS ElastiCache to host our Redis store. As a managed solution, all we had to do was define the specific configurations for the Redis cluster, such as port numbers, number of shards, numbers of nodes per shard, subnets and instance type. After the configuration was created, AWS ElastiCache automatically provisioned the Redis cluster for us.

We decided to use a single-sharded Redis cluster so we wouldn’t have to worry about equally distributing the key-value pairs among multiple shards. A simple technique like a modulo function on the number of shards would have worked, but it would be challenging to scale up the number of shards in the future. For example, if a cluster had 10 shards and the algorithm was “\% 10”:
\begin{itemize}
\item Key 309 would be stored in shard 9
\item Key 85 would be stored in shard 5
\item Key 234 would be stored in shard 4.
\end{itemize}
However, if there was a write-heavy workload and the number of shards had to be increased by 1 to 11, most keys in the cluster would have to be re-mapped to a different shard.
\begin{itemize}
\item Key 309 would be re-mapped to shard 1
\item Key 85 would be re-mapped to shard 8
\item Key 234 would be re-mapped to shard 3
\end{itemize}
Consistent hashing is a technique that would reduce the re-mappings required when the number of shards changes, but it’s complex and out of the scope of this project, so we left it as a possible future implementation.

Even though we only configured a single shard for our Redis cluster, we have two nodes for that shard. Both of the nodes store the same data, but have different APIs. One of the nodes is a primary (write) node and the other is a (read) replica. The primary node can be both written and read to, while the replica is only used for read operations. As a client, our application only has to write to the primary node, and the replica will automatically sync with the primary node’s data. This allows our application to scale horizontally for large read-heavy workflows by increasing the number of replicas when needed. The tradeoff, however, would be a loss of consistency guarantees. Although the data syncing from the replica is done at frequent intervals, there is a possibility that a new key is saved to the primary node but hasn’t propagated yet to the replica, before that new key is fetched. For an application like ours, it is fine if there is no strong consistency guarantees because users can just refresh the page if there’s a key-miss. However, for a financial or government application where missing data can have severe consequences, it may be a good idea to consider alternative database solutions.

\subsection{AWS Virtual Private Cloud}
\begin{figure}[htbp]
\centerline{\includegraphics[width=0.5\textwidth]{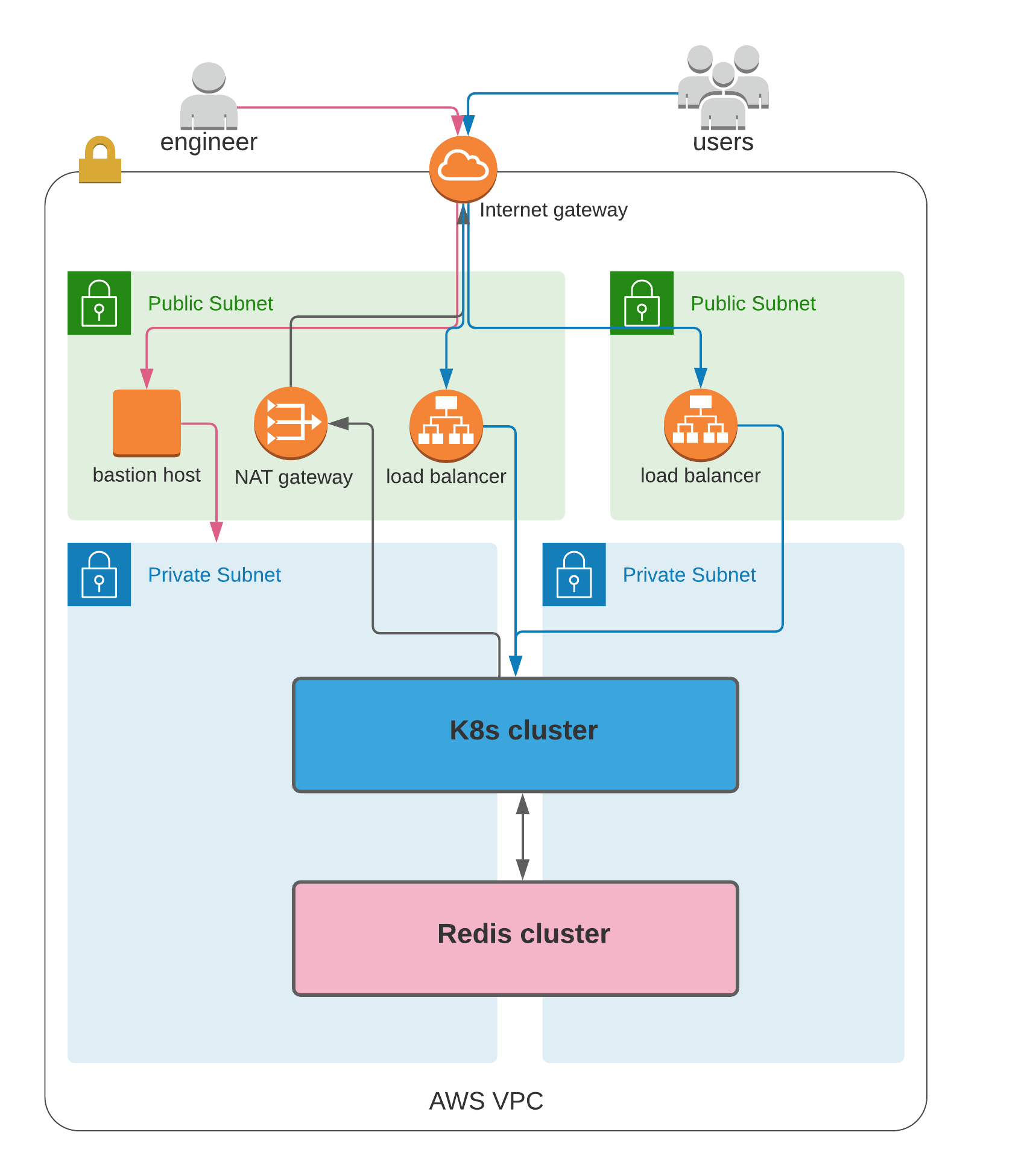}}
\caption{Note that the NAT gateway is only in a single public subnet. If its availability zone goes down, it means the application won't be able to initiate connections to the public Internet, so something like the Python "requests" module will stop working. However, the application will still continue to function.}
\label{fig:networking}
\end{figure}

In our application, we have 2 public subnets and 2 private subnets. The public subnets contain our proxy load balancers and bastion host, and the private subnets contain the instances hosting our Kubernetes and Redis clusters. The reason why we need 2 public and private subnets as opposed to 1 public and private subnet, is for fault tolerance. The limitation of AWS VPC is that instances within a subnet can only be spread within a single availability zone, which can be thought of as a physical data center. If an AWS data center goes down, we should still try to keep our application available. By duplicating proxy servers and cluster nodes across two separate availability zones/subnets, we prevent application downtime from issues in a single data center, such as a local power outage. Figure \ref{fig:networking} shows a diagram of the overall cloud architecture with the multiple subnets.

\subsection{Bastion Host}
Since our application servers were in private subnets, it was not possible to directly SSH into them for debugging any issues. We created a bastion host, which is a secure server that sits in a public subnet. The bastion host only has port 22 open for SSH, and all of its other ports are closed. Since the bastion host is within the same virtual network (VPC) as the application servers, it is possible to use the bastion host as a jump server to then SSH into an application server. First an engineer would SSH into the bastion host with its public IP address, and then SSH into any application server using the server’s private IP address.

In our class project demo, this did not work, with the application server rejecting the SSH request from the bastion host. This is because we did not SSH into the bastion host with \href{https://www.ssh.com/ssh/agent#ssh-agent-forwarding}{SSH agent forwarding} enabled. The bastion host does not and should not contain any SSH keys of the application servers, instead the SSH keychain of the engineer should temporarily be propagated to the bastion host, so that it could be reused for SSHing into the application server. To forward the SSH keychain when SSHing into the bastion host, the ssh command option “ForwardAgent” should have been set to “yes”.

\subsection{AWS Cloudformation}
To codify our major cloud resources, including the AWS VPC, AWS EKS cluster, AWS ElastiCache cluster, bastion host and security groups, we used \href{https://aws.amazon.com/cloudformation/}{AWS Cloudformation}. AWS Cloudformation is an IaC tool created specifically for provisioning AWS resources. We wrote YAML configuration files with the resource specifications we wanted, and automatically deployed the resources in those configuration files using the \href{https://aws.amazon.com/cli/}{AWS CLI} tool. AWS Cloudformation allowed us to keep track of what resources we created in our cloud environment, so we could easily modify and de-provision them later when necessary.

\subsection{Continuous deployment with Github Actions}
We used Github Actions for the continuous deployment (CD) pipeline. Github Actions is a service provided by Github, that allows engineers to write custom commands and scripts that are triggered after a commit is pushed to the hosted repository. Github Actions are structured in the form of workflows, jobs and steps. Somewhat analogous to Ansible, workflows are similar to playbooks, jobs are similar to tasks that can contain multiple steps. One large difference though, is that jobs run in parallel by default in a workflow, while in Ansible, tasks run sequentially. For a particular job, all of the steps are guaranteed to be executed on a single machine/deploy agent, so it is possible to install dependencies in one job step and use those dependencies in the next step.

After a commit is pushed to a repository, a newly created job is initially queued, before it is picked up by an agent. It is likely that Github as a company, allocates many machines to serve as agents for the purpose of running jobs. Each agent would primarily have two states, “available” and “busy”. Once a busy agent completes a job, its state then turns available until it picks up the next job off the globally distributed queue. There’s no way of exactly knowing how Github architects their Github Actions service though, without being employed at the company.

For our project, we have two different Github Action workflows. One of the workflows is responsible for building newly changed images and pushing them to the public Dockerhub registry. The other workflow checks if there is any change to the Kubernetes cluster configuration files, and if there are, applies those changes with the kubectl client.

Both of the workflows are dynamically triggered. Only when files in the \href{https://github.com/kevjin/rce-research/tree/master/images}{images/} directory are changed does the \texttt{engine-docker-images} workflow run, and it examines the changed files using the \href{https://github.com/kevjin/rce-research/blob/master/build_images.sh}{build\_images.sh} bash script to determine which image to re-build and push. The \texttt{k8s-cluster-deployment} workflow only runs when a configuration file in the \href{https://github.com/kevjin/rce-research/tree/master/kubernetes}{kubernetes/} directory is changed, and using the script \href{https://github.com/kevjin/rce-research/blob/master/apply_k8s_changes.sh}{apply\_k8s\_changes.sh}, it checks which configuration files were changed so that kubectl knows to only “apply” those files.

In order for the Github Actions agents to be able to access the Dockerhub registry and remote Kubernetes cluster hosted on AWS, it was necessary for us to provide those agents with the secrets to do so. Repositories on Github have a section in the Settings tab called “Secrets” that is only available to project administrators. The secrets we defined there, such as Docker username/passwords and AWS tokens, would be made available to the deploy agent automatically.

\subsection{Predictive Horizontal Pod Autoscaler}

We used an open-source tool called \href{https://predictive-horizontal-pod-autoscaler.readthedocs.io/en/latest/}{Predictive Horizontal Pod Autoscaler (PHPA)}, that is built on top of Kubernetes' native (reactive) \href{https://kubernetes.io/docs/tasks/run-application/horizontal-pod-autoscale/}{Horizontal Pod Autoscaler (HPA)}. HPA queries the Kubernetes’ API every 15 seconds by default to check if a specific service’s pods reach a configured CPU threshold \cite{HPA}. We specified 75\% as the CPU threshold. When the CPU threshold is hit, HPA automatically recommends additional pods up to a set limit (we specified 10 pods max). This way, if a particular application user is running a CPU-heavy experiment on the Python runlang service, new pods will be spun up and it won't affect other Python service users nearly as much. PHPA caches the historical recommendations by HPA, in order to forecast the optimal \# of pod replicas to provision for the next 10 seconds. In our implementation, we cache the last twelve replica count recommendations by HPA, in the last two minutes, in order to make a single-step forecast on what the pod replica count should be 10 seconds from now.

\begin{figure}[htbp]
\centerline{\includegraphics[width=0.5\textwidth]{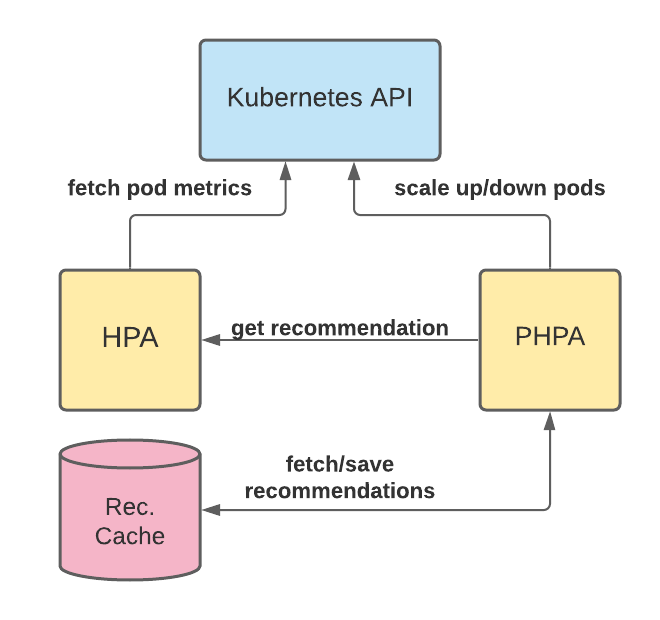}}
\caption{Pod scaling prediction pipeline}
\label{phpa-arch}
\end{figure}

We decided to use ten seconds as the interval between each data point and our prediction, because it roughly takes ten seconds maximum to spin up a pod in our experience. We had to choose a time that gave the new pods enough time to spin up, and not predict too far in the future, else it wouldn't be as useful for fast-changing traffic workloads. We experimented with two different machine-learning (ML) algorithms in order to determine which one works the best for our use case, and the results \& experiment process are described in detail in the section below.

\section{Experiment Set-up}
In order to determine which ML algorithms work best for scaling the remote code execution service, we decided to run several independent experiments comparing KNN and Linear Regression algorithms. We chose both of these models because they're simple and didn't require much data to make predictions, which is ideal given that we're only caching twelve historical data points.

Before we could run the experiments, however, we had to construct the test environment and build the following tools:
\begin{itemize}
\item World Cup 1998 Internet traffic data parser (Python)
\item Asynchronous load testing tool
\item Predictive autoscaler with select ML models
\end{itemize}

The goal was to have the traffic data parser tool read real, historical Internet traffic data (from the World Cup 1998 event) into the load testing tool, so that our remote code execution service could experience similar Internet traffic load patterns as the World Cup website. The load testing tool would record the latency time of all of its outgoing requests, or if the request failed. The predictive autoscaler was in charge of scaling up the pod replicas to hopefully minimize the recorded latencies.

\subsection{Internet Traffic Data Python Parser}
We decided to use the World Cup 1998 Internet Traffic data, as opposed to synthetic or random data because of various problems with the latter two. It is pointless to model and run forecasting algorithms on randomly generated traffic data, so we rejected the idea. Synthetic data is better, but it is difficult to generate good synthetic data that has similar patterns to real Internet traffic. The World Cup 1998 Internet Traffic data is a popular open-source dataset that has been used in many research studies, so we decided on it. It is a very large dataset with a history of all the incoming client requests between specific dates of the event. Unfortunately, it is from two decades ago, but general Internet traffic patterns shouldn't have changed enormously. There was a provided C parser for the raw byte data, but it wasn't suitable for our project that was primarily written in Python, so we re-wrote the data parser for our convenience.

\subsection{Asynchronous Load Testing Tool}
With the traffic data streaming in from the data parser, we needed a way to mimic the Internet traffic hitting our remote code execution service. What we noticed immediately was that Python's popular \emph{requests} library only supported synchronous requests \cite{requests}, which was unfeasible to fake our Internet traffic data with, that had at least multiple requests per second. We had to make some adjustments and built a load testing tool that batched all the HTTP requests in a five second period and asynchronously fired them together to our service.
Each HTTP request that the load testing tool sent was the same, for simplicity. It was a POST request to an endpoint that would trigger a short Python execution of the statement: \emph{print("hello there")}. Given enough concurrent requests, the service should scale up the number of ``runpython'' pod replicas.
The load testing tool was also responsible for keeping track of the response time of all of its outgoing requests or checking if the request failed. At the end of the load testing session, the tool would log detailed information regarding response times and the failed requests count.

\subsection{Predictive Autoscaler}
We found an open-source project on Github, \emph{predictive-horizontal-pod-autoscaler} (PHPA) \cite{PHPA}, that supported scaling up/down pod replicas according to forecasts by a selected machine learning model. Unfortunately, PHPA didn't support forecasting with the KNN model yet, so we forked and modified the project to implement it ourselves. We also added a control group model that would simply forecast the number of pod replicas needed to be the latest recommendation by Kubernetes' (reactive) HPA.

\subsection{Penalty Equation}
In order to determine each model's effectiveness in predicting the optimal number of runpython pod replicas, we derived a penalty equation that factors in:
\begin{itemize}
\item response time
\item total number of successful and failed requests
\item average number of pod replicas
\end{itemize}

For the remote code execution service, it is more detrimental to user experience when a single request takes ten seconds vs. two requests taking five seconds each. In order to more harshly penalize longer requests, we squared the average response time (of a single batch job) in our equation. Ideally, we would square the response time of each individual request instead, but that was not possible because of the asynchronous request limitations of the load testing tool.

\begin{align*}
  \frac{\sum_{i=0}^{\#\ of\ batches} \substack{successful\ \\ request\ count(i)} * \substack{average\ response\ \\ time(i)^2}}
        {\sum_{i=0}^{\#\ of\ batches} \substack{successful\\ \ request\ count(i)}} \\
\end{align*}

What is worse than a long request is a failed request. We penalize these failed requests as if they had a half-minute latency.
\[ failed\ request\ count * 30^2 \]

We also penalize the autoscaler if it over-provisioned the number of pod replicas. Having too many pod replicas in a cluster would be a waste of resources. We represent the average number of pod replicas throughout the experiment as the following variable in our cost equation:
\[ average\ pod\ replica\ count \]

Below is the final equation that brings all the above parts together.
\begin{align*}
  \frac{\sum_{i=0}^{\#\ of\ batches} \substack{successful\ \\ request\ count(i)} * \substack{average\ response\ \\ time(i)^2}}
        {\sum_{i=0}^{\#\ of\ batches} \substack{successful\\ \ request\ count(i)}} \\
    + failed\ request\ count * 30^2 \\
    + average\ pod\ replica\ count
\end{align*}

\section{Results}
We ran each experiment for 30 minutes with the selected ML model for predicting the optimal runpython pod replica count. Each experiment was performed twice to improve precision of the models' penalty scores.

\begin{figure}[htbp]
\centerline{\includegraphics[width=0.5\textwidth]{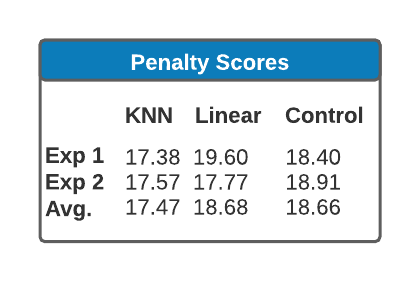}}
\caption{Penalty scores from all six experiment runs}
\label{fig:penalty}
\end{figure}

\begin{figure}[htbp]
\centerline{\includegraphics[width=0.5\textwidth]{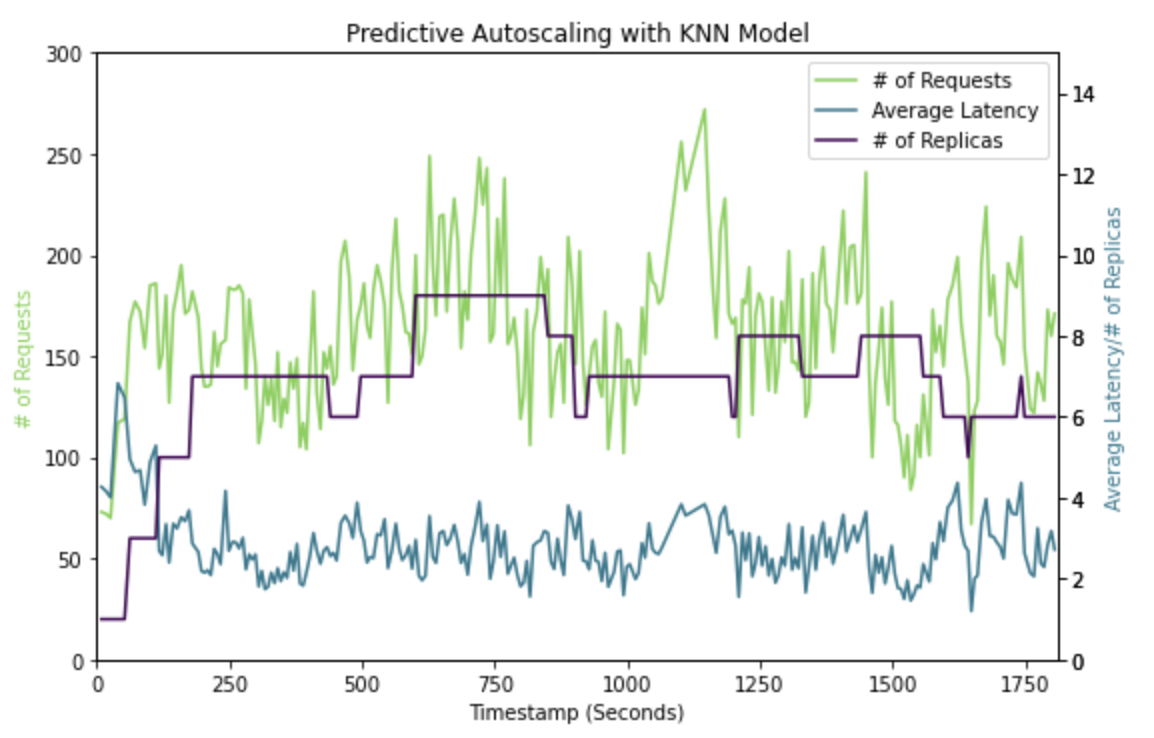}}
\caption{\# of failed requests: 77; Penalty equation score: 17.38}
\label{fig:knn}
\end{figure}

\begin{figure}[htbp]
\centerline{\includegraphics[width=0.5\textwidth]{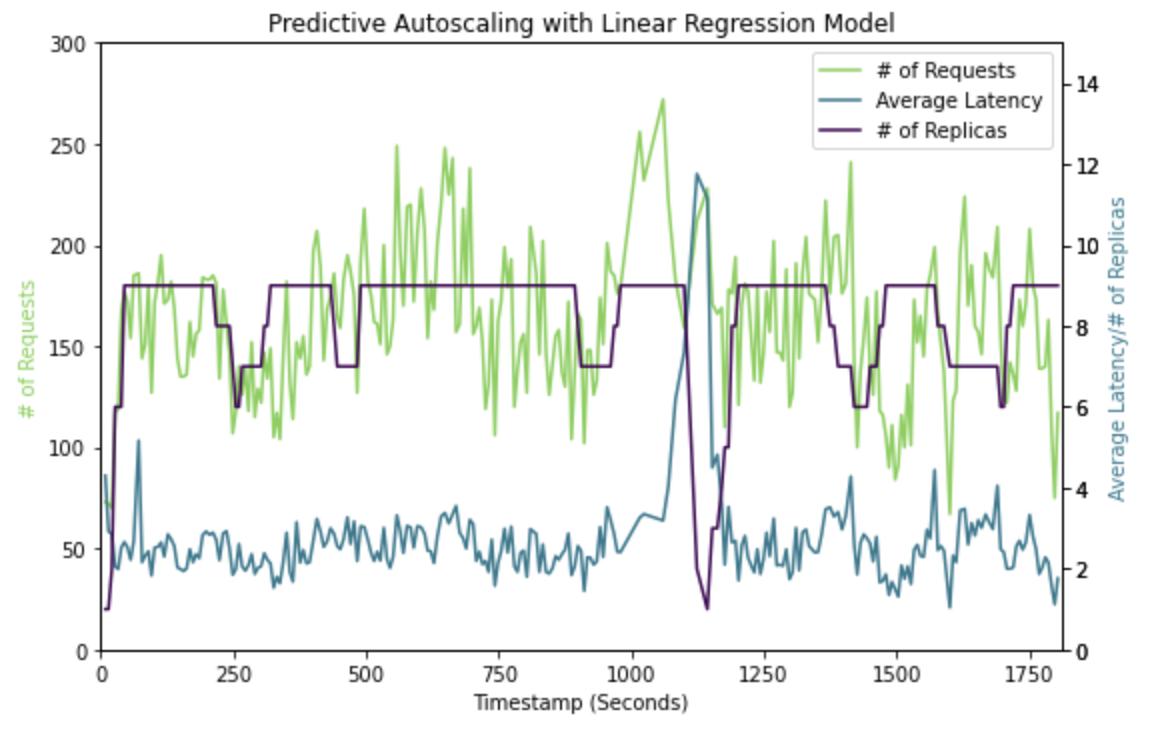}}
\caption{\# of failed requests: 123; Penalty equation score: 19.60}
\label{fig:linear}
\end{figure}

\begin{figure}[htbp]
\centerline{\includegraphics[width=0.5\textwidth]{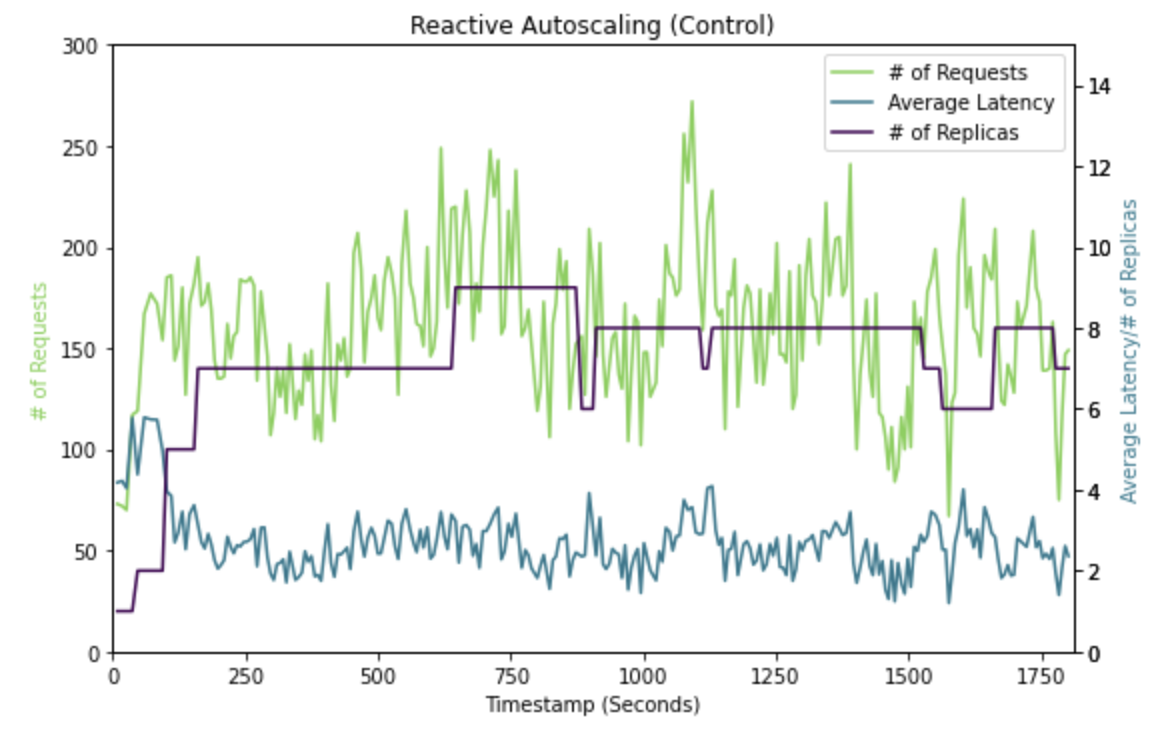}}
\caption{\# of failed requests: 65; Penalty equation score: 18.40}
\label{fig:reactive}
\end{figure}

Displayed in Fig \ref{fig:knn}, \ref{fig:linear} and \ref{fig:reactive} are the timestamped result data generated from the first experiment run of each model.

Notice that the \emph{\# of Requests} (green line) has a similar pattern in all of the experiments. This is because each experiment began at the same time on the 25th day (May 20, 1998) of the World Cup. Minor variability of the \emph{\# of Requests} is visible and expected because sometimes the batch asynchronous requests would take longer than the five second interval to finish, slightly delaying the next batch.

\section{Discussion}
Looking at the results in Fig \ref{fig:penalty}, we found that the KNN model consistently performed the best according to our initial penalty equation. Surprisingly, the reactive autoscaling actually ended up being slightly ahead of the Linear Regression model in terms of score, due to the underwhelming first run that the Linear Regression model had.

As shown in Fig \ref{fig:linear}, there was a large and unexpected plummet in the \# of Replicas in the first Linear Regression experiment around the 1200 seconds timestamp. That large drop in replica count also appeared to trigger an upwards spike in average latency, and definitely contributed to the high penalty score. We are not sure why the drop occurred, but it goes to show that predictive autoscalers may possibly overshoot their predictions and run into consequences.

Given how close the penalty scores were for the linear regression and reactive autoscalers, it would be interesting to try running the experiments several more times and see if the linear regression model ends up performning better than the control. Based on the results that we've collected so far, we can say that the KNN model performs better for our use case than both the linear regression model and control.

Taking a step back and looking at the penalty equation we used in the experiments, it's important to note that some values in it are slightly arbitrary. We penalize failed requests as if they had a 30 second latency, and we implicitly multiply the average pod replica count cost by a factor of 1. However, depending on the business requirements, it could be that failed requests are a SLA violation and should be penalized much more. Depending on the project budget, cloud resources could be more scarce and the pod replica count should be multiplied by a factor of 2-4. Changing these parameters in the penalty equation could also affect which predictive model performs best.

\begin{align*}
  \frac{\sum_{i=0}^{\#\ of\ batches} \substack{successful\ \\ request\ count(i)} * \substack{average\ response\ \\ time(i)^2}}
        {\sum_{i=0}^{\#\ of\ batches} \substack{successful\\ \ request\ count(i)}} \\
    + failed\ request\ count * \substack{FAILED\ REQUEST\ PENALTY} \\
    + average\ pod\ replica\ count * \substack{CLOUD\ RESOURCE\ \\ SCARCITY\ FACTOR}
\end{align*}

\section{Conclusion}

We have presented a web service for remote code execution, scale-able from small to very large workloads and easily extensible to support arbitrary language/ecosystem package configurations. By virtue of a modular design, our service scales easily without resource wastage and executes user code with minimal overhead.

Our service fills the niche of a platform supporting software experiments with a variety of programming languages and packages in those languages' ecosystems with zero start-up and tear-down cost incurred by the user. Such platforms are increasingly important in industrial software engineering environments, where experimentation of various software alternatives is necessary for software design, and the automation of provisioning one-time environments presents a significant resource/time cost reduction opportunity for businesses.
The service permits sharing of reproducible experiments between users in a workplace,
increasing its value in collaborative environments.

\section{Appendix}

\subsection{Research Motivation}

This paper originally started out as an undergraduate class project in Professor Aniruddha Gokhale's Cloud Computing course (CS-4287) at Vanderbilt University. The following semester, Kevin extended this project to add predictive horizontal autoscaling, as part of an undergraduate research study with Professor Gokhale.

\subsection{Division of Work}

Ayaz was responsible for design of the playground UI, web servers, and code execution images.

Kevin was responsible for scaling the service with Kubernetes, managing deployments onto AWS, architecting a secure network to host the application, and running the predictive autoscaling experiments.

Note, this delineation of work was not absolute; team members did what was required of the
implementation plan to get the project done collaboratively.

\subsection{Development Platforms Used}

The service was developed in a \texttt{git} monorepo hosted on \href{https://github.com/kevjin/cs4287-final}{GitHub}.
Images for components of the service are available as public repositories on \href{https://hub.docker.com/u/ayazhafiz}{Dockerhub}.

CD pipelines were built to increase velocity and correctness of the development process.
\href{https://docs.github.com/en/free-pro-team@latest/actions}{Github Actions} is used to test the service's components and publish builds
of component images.

\subsection{Testing strategy}

Integration testing is done manually. Setting up a reproducible and automated integration testing
strategy would have been an endeavor larger than the project described by this report, and we felt
confident that our unit and manual tests would sufficiently validate the service behavior.

\subsection{Additional Collaborative Frameworks Used}

Google Docs and Overleaf were used for collaborative work on the project implementation plan and
final report.

\section{Future Work}

Our service lays the framework for a collaborative and rich software experimentation platform, and
there are open problems to solve not covered by the service design described in this report:

\begin{itemize}
    \item How can code execution be further sandboxed from container network/filesystem/process access
    by bad actors? Performing every code execution request in a separate container is too expensive
    to be considered seriously, but exposing a shared container environment between code execution
    requests opens the door to potential exploits. One approach is to use process-level control groups.
    \item How can ecosystem packages by loaded on a per-execution basis during a runlang container
    runtime? Loading a fresh build of packages for each execution request is very costly, and caching
    loaded package builds runs into issues of versioning that are not trivial to solve. Furthermore,
    how can loaded packages be audited for malicious intent in an automated way? Both of these problems
    have been discussed in the industry and in academic literature, but they are still open problems
    that would require custom solutions for our use case.
    \item How can ``project''-level software experiments, composed of multiple files, be performed?
    Naive solutions like simply creating a new project with fresh dependencies for each code execution
    request do not scale. More advanced techniques like build caches and session-based container
    provisioning face correctness and consistency issues. Here, ``pre-composed'' cloud IDEs like 
    StackBlitz \cite{stackblitz} provide some interesting ideas proven to work at medium-to-large
    scales.
\end{itemize}

We have made available the source code for our service as a Github repository at \href{https://github.com/kevjin/rce-research}{https://github.com/kevjin/rce-research}.

\end{document}